\documentclass[journal]{IEEEtran}
\usepackage{array}
\usepackage{multirow}
\usepackage[table]{xcolor}
\usepackage{cite}
\usepackage{url}

\usepackage[pdftex]{graphicx}
\usepackage{epstopdf}

\graphicspath{{./images-greyscale/}}
 \DeclareGraphicsExtensions{.pdf,.jpeg,.png,.eps}

\usepackage[cmex10]{amsmath}
\usepackage[tight,footnotesize]{subfigure}
\usepackage[justification=centering]{caption}
\usepackage[protrusion=true,expansion=true]{microtype}

\begin{document}

\title{FPGA Implementation of a Scalable and Run-Time Adaptable Multi-Standard Packet Detector}

\author{	James~Chacko, Marko~Jacovic, Nagarajan Kandasamy, Kapil~R. Dandekar \\
		Department of Electrical and Computer Engineering \\
		Drexel University, 3141 Chestnut St, Philadelphia, PA, 19104\\
		\{jjc652, mj355, nk78, krd26\}@drexel.edu
}

\maketitle

\begin{abstract}
\boldmath 
This paper describes a step by step approach for implementing a scalable and run-time adaptable multi-standard packet detector for orthogonal frequency divisional multiplexing (OFDM) based communication standards. The paper briefly describes considerations and design choices in making a modular system block with generic control supporting rapid prototyping and implementation of preamble-based packet detectors. The results were generated through implementation on a Xilinx Virtex-6 FPGA with a MicroBlaze processor instantiated to provide run-time control and adaptability. 
\end{abstract}

\vspace{-5pt}
\section{Introduction}
\label{sec: intro}

\IEEEPARstart{R}{ecent} research in wireless communications is focused primarily on techniques involving Orthogonal Frequency Division Multiplexing (OFDM), a digital multi-carrier transmission scheme that is advantageous due to its spectral efficiency and resistance to severe channel conditions as described in \cite{Proakis}. Considering the scarcity of available spectrum and the importance of correcting a received wireless signal after it propagates through a channel, OFDM is a highly favorable scheme to implement, as is seen in its wide usage in wireless local area networks based on IEEE 802.11, WiMAX, Long Term Evolution, and various other applications. 

To recover a received OFDM signal, it is necessary to implement synchronization between the receiver and transmitter and correct for various impairments related to the channel and the Radio Frequency (RF) Front Ends. Synchronization consists of both frequency and time components. A receiver is described as coherent if it is synchronized in frequency with the transmitter, and non-coherent if it is not. Frequency synchronization is typically not always required, but improves system performance. OFDM is highly sensitive to frequency offsets, and performance may be vastly degraded if not synchronized. Timing synchronization is always required for any type of receiver so that the information data in the payload of a packet and training data that is required to correct impairments, including for the purpose of performing frequency synchronization, may be recovered. For a received OFDM packet, it is necessary to determine the starting and ending points of the training and data components. The focus of this paper is on the timing synchronization aspect of OFDM, which may be accomplished through a process known as packet detection.    

Packet detection is a computationally intensive procedure performed at the receiver that is generally split into phases involving energy, coarse, and fine detection routines. Received Signal Strength Indicator (RSSI), if available from the Analog to Digital converters (ADC) on the RF front-end, can also be utilized to help start the detection process. Energy detection other than the RSSI is based on calculating the average power of the incoming data, while coarse and fine detection are based on auto-correlation and cross-correlation routines, respectively. In this paper, we describe our steps in developing a flexible packet detection technique for a research platform used for SDR development across different OFDM based communication standards. 

The key differences in the implementation of multi-standard packet detectors are the variable length and value of the preamble that is used for detection. This functionality is achieved by the implementation of a variable point cross-correlation framework with correlation coefficients supplied as an input. The system design described in this paper is built in a modular way allowing the system to scale. In addition, the system was designed to take advantage of the increasing availability of logic slices, utilizing less DSP slices on the FPGA. The design is also run-time adaptable through a MicroBlaze instantiated on the Xilinx Virtex-6 FPGA to configure control parameters dynamically.

The structure of this paper is as follows: Section \ref{sec:backgrounds}  describes the theoretical background behind timing synchronization, Section \ref{sec:implementation} describes our implementation of a FPGA based scalable run-time adaptable packet detector explaining appropriate design considerations, Section \ref{sec:results} presents the simulation results and Section \ref{sec:conclusion} provides concluding remarks.
\begin{figure*}[t]
	 \includegraphics[trim=0mm 0mm 0mm 0mm, clip, width=6.2in]{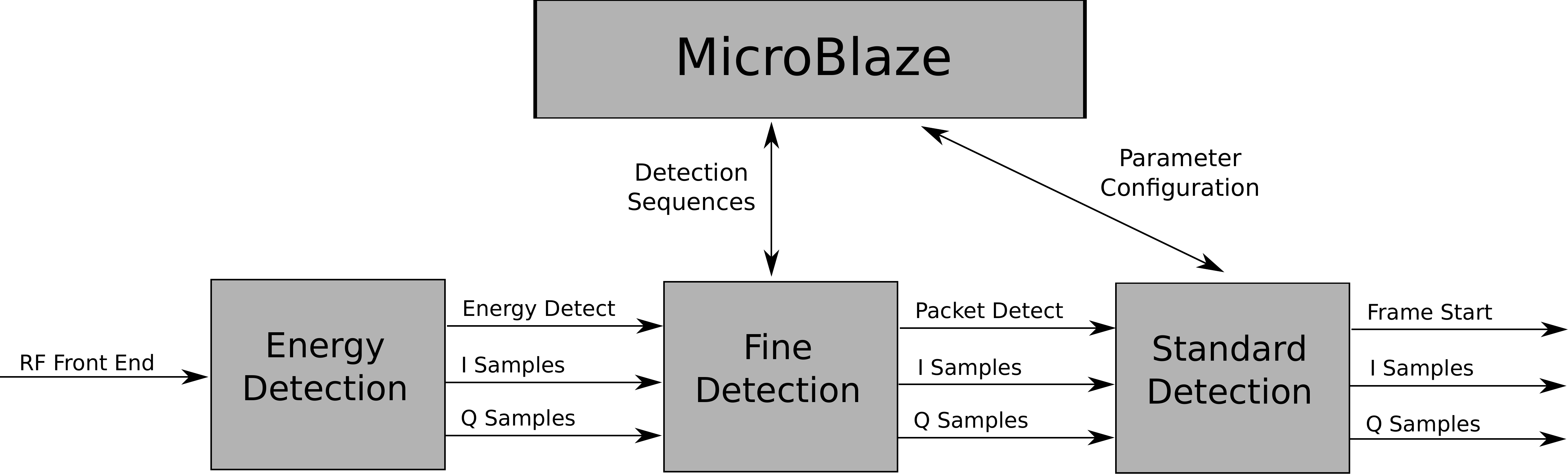}
	\centering
	\caption{Packet Detector design layout.}
	\label{fig: time}
\end{figure*}

\section{Background}\label{sec:backgrounds}
Our previous research in the implementation of a real time and protocol aware reactive jamming systems \cite{nguyen2014real} used the same principle described in this paper. However, in this paper we build upon these techniques to describe and implement a multi-standard protocol on a FPGA subject to specific constraints on modular scaling, control logic and resources. The implementation of cross-correlators on FPGAs for the purpose of timing synchronization has been done previously for IEEE802.11A in \cite{Gu} and \cite{Xie} and for WLAN in \cite{Wang}, but these cases are limited in the flexibility of the preambles used. Different preamble types have been used in \cite{Dick}, \cite{Tian}, and \cite{Qiang}, but in all cases there have been no consideration of parallel implementations for multiple signal types with different preamble structures. We are not aware of any other real-time FPGA implementation of multi-standard protocol detectors for SDR in the available literature. Based on the RF front-end being used, there may or may not be a RSSI value available at the receiver. For this reason, an energy-based detector was included as part of the design. A RSSI enabled front-end can always supplement the energy detector to reduce the occurrence of a false detection. 

The coarse detection step employs the Shimdl-Cox \cite{Schmidl} technique of auto-correlation, which sends a repeated sequence to be delayed and correlated with itself at the receiver to determine the start of the packet. This paper is focused primarily on fine detection which is based on cross-correlation and the layout of our design can be seen in Fig 1. The correlation coefficients used for this purpose are found within communication standards. The method used in this work consists of an energy detector and a matched filter, with the filter only activated if a decision is made by the energy detector that a signal is present. The energy of the signal can be calculated using the equation:
\begin{equation}
\label{eqn: energy}
E_{s} = \sum^{N}_{n=0} |{y[n]}| ^{2}
\end{equation}
where $y[n]$ and N are the received signal and observation window length, respectively. The implemented match filter can be represented as a cross-correlator described by
\begin{equation}
\label{eqn: corr}
P[n] = \sum^{\infty}_{m=-\infty} y^{*}[m]h[m+n]
\end{equation}
where $h[n]$ is the reference signal. As shown in \cite{milcom}, the in-phase and quadrature components of the cross-correlation can be simplified and written as
\begin{equation}
\label{eqn: hwcorr}
\begin{split}
Re\{P[n]\} = P_{II}[n]+P_{QQ}[n] \\
Im\{P[n]\} = P_{QI}[n]-P_{IQ}[n]
\end{split}
\end{equation}
where $P_{II}[n]$ is the cross-correlation of the in-phase component of the received signal and the in-phase component of the reference signal,  $P_{QQ}[n]$ is the cross-correlation of the quadrature component of the received signal and the quadrature component of the reference signal, $P_{QI}[n]$ is the cross-correlation of the quadrature component of the received signal and the in-phase component of the reference signal, and $P_{IQ}[n]$ is the cross-correlation of the in-phase and quadrature components of the received signal and reference signal respectively. The computation of both the in-phase and quadrature components are implemented through separate flows to handle complex arithmetic on hardware.

\section{Hardware Implementation} \label{sec:implementation} 
The system was designed mainly using System Generator on top of Drexel University's Software Defined Communication (SDC) \cite{SDC} testbed to target a Virtex-6 ML605 FPGA evaluation board. SDC is a highly scalable design platform that, unlike other available platforms, obtains its flexibility from its Scalable Orthogonal Frequency Division Multiplexing (SOFDM) core built specifically for rapid prototyping~\cite{chacko2015rapid} of OFDM-based baseband/physical (PHY) layer research \cite{SDC2}. The packet detector described in this paper shares its modular architectural development flow from SOFDM, allowing SDC to be used for multi-standard packet detection. The packet detector developed on hardware consists of \emph{Energy Detection}, \emph{Fine Detection}, and  \emph{Standard Detection} components. An overview of the development for each module is described below.
\vspace{1mm}
\subsubsection{\underline{Energy Detection}} 
The computation of \eqref{eqn: energy} is used to determine the presence of a signal by comparing the output value to a predetermined threshold. For time-smoothing purposes, addressable shift registers are used on the in-phase and quadrature samples that are considered in the calculation. The number of energy samples used in the computation is variable for the system. An energy detection event is determined if the number of energy samples within a window is greater than a configurable threshold determined over channel state.

\vspace{1mm}
\subsubsection{\underline{Fine Detection}} 
The decision from the energy detector is used as an enable signal for the fine detection, implemented through a cross-correlator, to start processing. Although it is possible to have the cross-correlator always running, it increases the possibility of a false detection and is power inefficient. To make a multi-standard packet detector, the two main aspects considered were to design the correlation coefficients as variables into the cross-correlator and to build modularity in a manner that extends functionality for longer cross-correlations with ease through stacking. 

Aside from the enable signal from the energy detector, the only other signal into the fine detection module is the received complex signal split into its in-phase (I) and quadrature (Q) components. Since computing a $n$-point correlation involves $n$ multiplications and $n$ additions according to \eqref{eqn: corr}, it is very expensive in terms of DSP slices. In order to reduce the complexity, the incoming I/Q signals are both sent through a categorizer module to convert the signals either to a positive or negative value of one. The reduction of the signal into its respective positive and negative units significantly reduces the complexity, as simple arithmetic operations of addition and subtraction are implemented instead of the $n$ multipliers.  

The correlation coefficients are variables to the system set through the MicroBlaze processor written into a 32-bit shared register and unique in being able to extend on demand unlike other implementations. The register value is generated using a function to map the known preamble through a threshold function that produces $0$ for a complex value below $0$ and $1$ for a complex value greater than or equal to $0$, after which the $0$s and $1$'s are bit-bashed to load the coefficient register. A set of unsigned 32-bit fixed-point registers are initialized for this purpose where each bit represents the sign of the known preamble for the incoming signal that it has to be compared against. To explain further, a 16-point correlation would only use the coefficient loaded onto half a register while a 32-point correlation would use data loaded onto the entire 32-bit register.
\begin{figure}[t]
	 \includegraphics[trim=0mm 0mm 0mm 0mm, clip, width=3.5in]{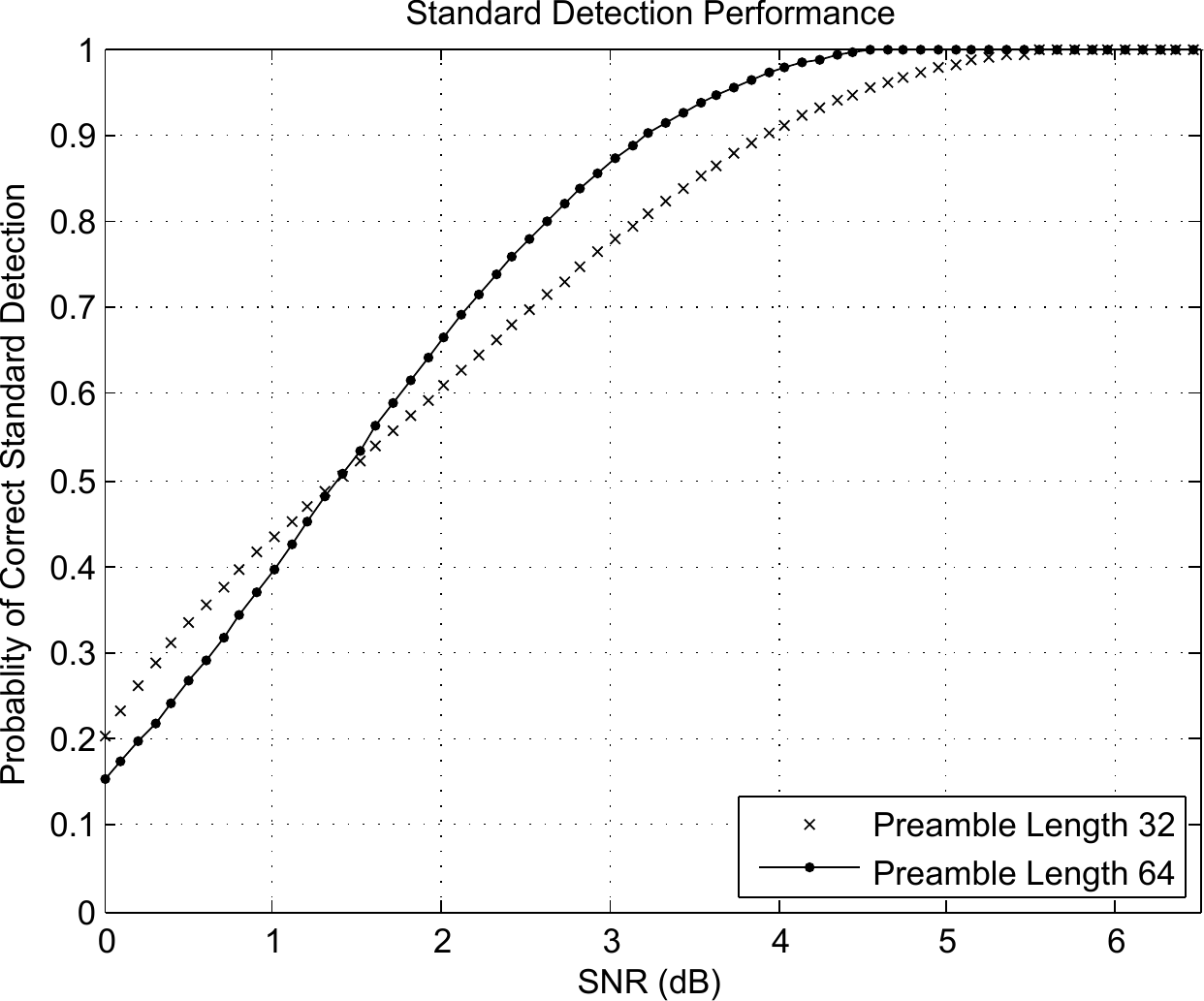}
	\caption{Hardware Simulation Detection Performance Results.}
	\label{fig: hwsim}
\end{figure}
\begin{figure}[t]
	 \includegraphics[trim=0mm 0mm 0mm 0mm, clip, width=3.5in, height=5in]{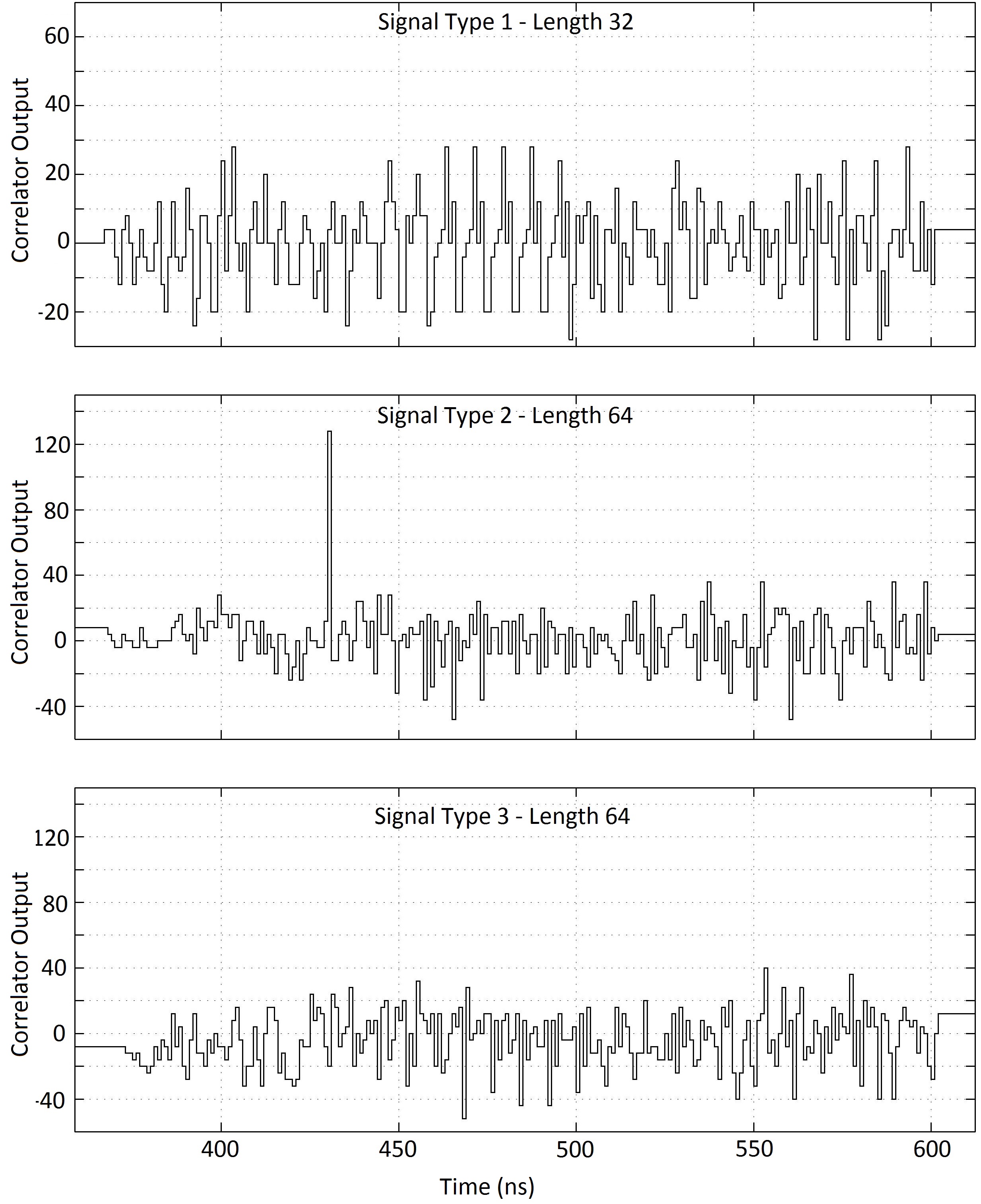}
	\caption{Hardware Simulation Cross-Correlator \\Output Comparison.}
	\label{fig: scope}
\end{figure}
The incoming data must be windowed in order to parallelize the computation of \eqref{eqn: corr}. To implement a scalable window size, it is important to design a system which buffers incoming data with the ability to have values extracted across the stored values. Although varying delay blocks would have seemed to be the fastest way to implement the windowing system, the method would have been more expensive in terms of logic slices on the FPGA. Instead, variable shift registers are used to implement the buffering and to allow all data within the window be used with the stored coefficient in computation. To allow for expansion, the windowing cores can be stacked in parallel and accept parameters.
\vspace{1mm}
\subsubsection{\underline{Standard Detection}} 
The standard detection block enables the appropriate parameters for the detected standard. These parameters includes packet length, symbol sizes, training period, and other relevant settings for retrieving original data. When this design is targeted on FPGAs with on-board processors, such as the MicroBlaze, it takes an active role in this stage of packet detection by loading the shared parameter registers instead of having the values set through relational MUXs or equivalent logic. 

\section{Simulation Results} \label{sec:results}
Selecting a preamble with desirable auto-correlation function attributes, that have a high peak correlation value with a delay of zero and very low output values at the sidebands, may be preferable for the purpose of timing synchronization. A common consideration for pseudo-noise preambles based on this desired characteristic is a Barker sequence as described in \cite{barker}. Barker sequences are limited in that only a single unique sequence exists at lengths of 11 and 13 samples long. Shorter Barker sequences were not considered in this study as their usage would be prone higher to false detection. For this reason, this study was performed by selecting a randomly generated sequence to highlight the high flexibility of the system and demonstrate how it is not restricted to single standard specific preambles. The design is capable of accepting any pseudo-noise reference signal, and may easily be adjusted for preamble length if a specific signal, such as one with favorable auto-correlation function features, is desired. 

To observe the performance of the design, the system was configured to compare three arbitrary signal types in parallel with different randomly generated complex pseudo-noise preambles. The first signal was chosen to have a length of 32 samples, while the other two had lengths of 64 samples. The input signal to the hardware simulation was generated in MATLAB script by arbitrarily selecting the second preamble type to be used as the signal preamble, adding a desired amount of noise to the signal, and separating the signal into real and imaginary components. For a complex preamble, the ideal maximum value of a 32-point correlator is 64, while for a 64-point correlator the ideal maximum value is 128. Due to the presence of noise, it is necessary to soften the correlator decision threshold for detection, with values of 50 and 100 chosen respectively in this study for the two sample lengths. In the event that both the 32 and 64 sample long preambles were detected, priority is given to the 64 sample preamble detection because it is less probable to occur as a false alarm. A scope capture from the hardware simulation for the described scenario with a Signal to Noise Ratio (SNR) of 10 dB is shown in Fig \ref{fig: scope}. A comparison of the correlator outputs shows that the correct signal was detected in the hardware simulation using the second preamble. The second plot of the scope capture yielded a peak well above the threshold of 100, while in the first and third plots the correlator outputs do not reach the necessary thresholds of 50 and 100. These results verify the capability of the system to perform simultaneous processing in time to compare multiple preambles known to the receiver to the received signal. Timing synchronization may then be implemented by delaying the received samples based on the location of the peak correlation output. 

The system performance was examined over a larger range of SNR values for both 32 and 64 sample length preambles. In each scenario the signal types previously described were used for comparison at the receiver. The probability of correct detection was approximated through post-processing of the correct number of signal type decisions for each respective preamble length over 300 repeated trials at each SNR value. In this study, both missed detections and false alarms of other signal types were treated identically. The study was performed in two separate stages to analyze performance: first selecting a preamble type of 32 samples long and comparing it at the receiver to the two unique 64 sample long preambles, then selecting a signal type of 64 sample preamble length and comparing it to the 32 sample preamble type and the other unique 64 sample preamble. As seen in Fig \ref{fig: hwsim}, the 64 sample long preamble outperforms the 32 sample long type. This may be attributed to the auto-correlation function of the 64 sample long preamble having a higher ratio between the main and sidelobes of the correlation output in comparison to the 32 sample long preamble. However, it is important to note that both perform well overall in low SNR conditions, and would be capable of being used successfully in packet detection. 

\section{Conclusion} \label{sec:conclusion}
In this paper a step by step approach in building a scalable and modular packet detector capable of detecting packets across multiple communication standards for wireless OFDM was described. The focus in this paper was on the implementation of a timing synchronization scheme consisting of energy detection, fine detection, and standard detection components. Results generated in this paper are based off of a Xilinx ML605 FPGA board design, showing successful packet detection across different preamble header scenarios.

\section*{Acknowledgment}\label{sec:acknowledgment}
This project is supported by the National Science Foundation through grants 1457306 and 1422964. 

\bibliographystyle{IEEEtran}
\bibliography{./chacko_bib_v3}

\begin{thebibliography}{10}
\providecommand{\url}[1]{#1}
\csname url@samestyle\endcsname
\providecommand{\newblock}{\relax}
\providecommand{\bibinfo}[2]{#2}
\providecommand{\BIBentrySTDinterwordspacing}{\spaceskip=0pt\relax}
\providecommand{\BIBentryALTinterwordstretchfactor}{4}
\providecommand{\BIBentryALTinterwordspacing}{\spaceskip=\fontdimen2\font plus
\BIBentryALTinterwordstretchfactor\fontdimen3\font minus
  \fontdimen4\font\relax}
\providecommand{\BIBforeignlanguage}[2]{{%
\expandafter\ifx\csname l@#1\endcsname\relax
\typeout{** WARNING: IEEEtran.bst: No hyphenation pattern has been}%
\typeout{** loaded for the language `#1'. Using the pattern for}%
\typeout{** the default language instead.}%
\else
\language=\csname l@#1\endcsname
\fi
#2}}
\providecommand{\BIBdecl}{\relax}
\BIBdecl

\bibitem{Proakis}
J.~G. Proakis and M.~Salehi, \emph{Digital communications}, 5th~ed.\hskip 1em
  plus 0.5em minus 0.4em\relax Boston, Mass: McGraw-Hill, 2007.

\bibitem{nguyen2014real}
D.~Nguyen, C.~Sahin, B.~Shishkin, N.~Kandasamy, and K.~R. Dandekar, ``A
  real-time and protocol-aware reactive jamming framework built on
  software-defined radios,'' in \emph{Proc. of {ACM SIGCOMM} Software Radio
  Implementation Forum ({SRIF '14})}, 2014.

\bibitem{Gu}
Y.~Gu, Y.~Li, W.~Ren, and K.~Chen, ``{FPGA} implementation of the timing
  synchronization in preamble based {OFDM} system,'' in \emph{International
  Conf. on Mobile Technology, Applications and Systems}, Nov 2005, pp. 4
  pp.--4.

\bibitem{Xie}
J.~Xie, Y.~Ding, S.~Yang, and L.~Qi, ``{FPGA} implementation of frame
  synchronization and symbol timing synchronization based on {OFDM} system for
  {IEEE} 802.11a,'' in \emph{Intelligent Signal Processing and Communication
  Systems (ISPACS), 2010 International Symposium on}, Dec 2010, pp. 1--4.

\bibitem{Wang}
K.~Wang, J.~Singh, and M.~Faulkner, ``{FPGA} implementation of an {OFDM}-{WLAN}
  synchronizer,'' in \emph{Proceedings of IEEE International Conf. on
  Field-Programmable Technology.}, Jan 2004, pp. 89--94.

\bibitem{Dick}
C.~Dick and F.~Harris, ``{FPGA} implementation of an {OFDM} {PHY},'' in
  \emph{Thirty-Seventh Asilomar Conference on Signals, Systems and Computers,
  2004.}, vol.~1, Nov 2003, pp. 905--909 Vol.1.

\bibitem{Tian}
Y.~Tian, R.~Wang, Q.~Li, L.~Liu, and S.~Gao, ``Design and implementation of
  {OFDM} timing synchronization with repeated-structured training sequence
  based on {FPGA},'' in \emph{Software Engineering and Service Science
  (ICSESS), 2014 5th IEEE International Conference on}, June 2014, pp.
  1099--1102.

\bibitem{Qiang}
W.~Qiang, T.~Cheng, and H.~Wei, ``Efficient implementation of synchronization
  in {OFDM} system based on {FPGA},'' in \emph{International Conference on
  Advanced Communication Technology}, vol.~1, Feb 2007, pp. 178--181.

\bibitem{Schmidl}
T.~Schmidl and D.~Cox, ``Robust frequency and timing synchronization for
  {OFDM},'' \emph{Communications, IEEE Transactions on}, vol.~45, no.~12, pp.
  1613--1621, Dec 1997.

\bibitem{milcom}
M.~Jacovic, J.~Chacko, D.~Pfeil, N.~Kandasamy, and K.~R. Dandekar, ``Hardware
  implementation of low-overhead data aided timing and carrier frequency offset
  correction for ofdm signals,'' in \emph{Military Communications Conference,
  MILCOM 2015 - 2015 IEEE}, Oct 2015, pp. 495--500.

\bibitem{SDC}
B.~Shishkin, D.~Pfeil, D.~Nguyen, K.~Wanuga, J.~Chacko, J.~Johnson,
  N.~Kandasamy, T.~Kurzweg, and K.~Dandekar, ``{SDC} testbed: Software defined
  communications testbed for wireless radio and optical networking,'' in
  \emph{Modeling and Optimization in Mobile, Ad Hoc and Wireless Networks
  (WiOpt), 2011 International Symposium on}, May 2011, pp. 300--306.

\bibitem{chacko2015rapid}
J.~Chacko, C.~Sahin, D.~Pfeil, N.~Kandasamy, and K.~Dandekar, ``Rapid
  prototyping of wireless physical layer modules using flexible
  software/hardware design flow,'' in \emph{{Proceedings of the 2015 ACM/SIGDA
  International Symposium on Field-Programmable Gate Arrays (FPGA '15)}}.\hskip
  1em plus 0.5em minus 0.4em\relax ACM, 2015, pp. 32--35.

\bibitem{SDC2}
J.~Chacko, C.~Sahin, D.~Nguyen, D.~Pfeil, N.~Kandasamy, and K.~Dandekar,
  ``{FPGA}-based latency-insensitive {OFDM} pipeline for wireless research,''
  in \emph{High Performance Extreme Computing Conference (HPEC), 2014 IEEE},
  Sept 2014, pp. 1--6.

\bibitem{barker}
S.~J. Lee and J.~Ahn, ``Acquisition performance improvement by barker sequence
  repetition in a preamble for ds-cdma systems with symbol-length spreading
  codes,'' \emph{Vehicular Technology, IEEE Transactions on}, vol.~52, no.~1,
  pp. 127--131, Jan 2003.

\end{thebibliography}

\end{document}